\documentclass[twocolumn,eqsecnum,aps,prb,showpacs]{revtex4}

\usepackage{graphicx}

\begin{document}   

\title{Electromagnetic response and effective gauge theory of \\ 
graphene in a magnetic field}
\author{K. Shizuya}
\affiliation{Yukawa Institute for Theoretical Physics\\
Kyoto University,~Kyoto 606-8502,~Japan }

\begin{abstract} 
The electromagnetic response of graphene in a magnetic field is studied,
with particular emphasis on the quantum features of its ground state
(vacuum).  The graphene vacuum, unlike in conventional quantum Hall systems,
is a dielectric medium and carries  an appreciable amount of electric and
magnetic susceptibilities. The dielectric effect grows rapidly with
increasing filling factor $\nu$ in such a way that reflects the
$\lq$relativistic' Landau-level characteristics of graphene as well as its
valley and spin degeneracy.  A close look into the dielectric function also
reveals that the Coulomb interaction is efficiently screened on the scale
of  the magnetic length, leading to a prominent reduction of the exciton
spectra in graphene.  In addition, an effective gauge theory of graphene is
constructed out of the response. It is pointed out thereby
that the electric susceptibility is generally expressed as a ratio of the
Hall conductance to the Landau gap.
\end{abstract}


\maketitle

\section{Introduction}

There has recently been increasing interest, both
experimentally~\cite{NG,ZTSK,ZJS} and
theoretically~\cite{ZA,GS,PGN,NM,AF,Massgap,FL}, in a
$\lq$relativistic' condensed-matter system,  graphene which is a
monolayer of graphite. Graphene is a gapless planar semiconductor, in
which low-energy electronic transport is essentially governed by massless
Dirac fermions with effective speed of light 
$v_{\rm F} \approx 10^{6}$ m/s $\approx c/300$, and thus provides a special
opportunity to study relativistic quantum dynamics in condensed-matter
systems. Experiments have revealed a number of exotic transport properties of
graphene, such as 
an unusual sequence~\cite{NG,ZTSK} of magneto-oscillations and of 
the quantum Hall (QH) effect, that are characteristic of Dirac fermions.

It has long been known that Dirac electrons 
in 2+1 dimensions lead to peculiar
quantum phenomena of fractional fermion number~\cite{J}:
Fractional charge and current of abnormal parity, as summarized by the
Chern-Simons (CS) term, are induced in the vacuum in response to an
applied field.
The induced CS term, or a parity anomaly, is associated with a spectral
asymmetry and an index of the Dirac Hamiltonian, 
and is also tied to the chiral anomaly in 1+1 dimensions~\cite{NS}.
The one-half degeneracy of the lowest Landau level
observed~\cite{NG,ZTSK} in the graphene QH effect is a manifestation of
fermion number fractionalization, although complete cancellation of the
net vacuum charge and current takes place in graphene (which involves 
a parity pair of fermions).

The purpose of this paper is to explore possible signatures of
relativistic quantum field theory in the low-energy physics of graphene.
We study the electromagnetic response of graphene in a
magnetic field at integer filling factor $\nu$, with particular emphasis
on the quantum features of the ground state with $\nu=0$, the vacuum
state. For graphene, unlike standard QH systems, the vacuum state is 
a dielectric medium and carries an appreciable amount of 
both electric and magnetic susceptibilities 
$(\alpha_{\rm e}, \alpha_{\rm m})$ over the entire range of wavelengths;
this reflects the presence of the $\lq$Dirac sea'.
The dielectric effect grows rapidly with
increasing filling factor $\nu$ in such a way that reflects 
the $\lq$relativistic' Landau-level characteristics of graphene
as well as its valley and spin degeneracy.
A close look into the dielectric function reveals that the Coulomb
interaction is efficiently screened on the scale of the magnetic length,
leading to a prominent reduction of the exciton spectra in graphene.

In addition, we construct, out of the response and via functional
bosonization, a low-energy effective gauge theory of graphene in a magnetic
field. It is pointed out thereby that the electric  susceptibility
$\alpha_{\rm e}$ is generally expressed as a ratio of the Hall conductance
to the Landau gap.

In Sec.~II we study the electromagnetic response of graphene 
in a magnetic field, with emphasis on  projection to Landau
levels and regularization.
In Sec.~III we construct an effective theory of graphene.
In Sec.~IV we discuss the effect of polarization by treating the Coulomb
interaction in the random-phase approximation (RPA).
Section~V is devoted to a summary and discussion.
\\

\section{Electromagnetic response}

Graphene has a honeycomb lattice which consists of two triangle
sublattices of carbon atoms with one electron per site.
The electrons are described by a two-component spinor $(U,V)^{\rm t}$
with fields $U$ and $V$ residing on each sublattice.
In the tight-binding approximation with nearest neighbor interactions
the spectrum of electrons becomes linear at the two inequivalent Fermi
points ($K$ and $K'$) on the corners of the Brillouin zone.
In the continuum limit the system is described by
a low-energy effective Hamiltonian of the form~\cite{Semenoff}
\begin{eqnarray}
H&=& \int d^{2}{\bf x}\Big[ \psi^{\dag} {\cal H}_{+} \psi 
+ \chi^{\dag} {\cal H}_{-}\chi \Big], \nonumber\\
{\cal H}_{\pm}&=& v_{\rm F}\,  (\sigma_{1}\Pi_{1} 
+ \sigma_{2}\Pi_{2} \pm m \sigma_{3})  - e A_{0}, 
\label{Hzero}
\end{eqnarray}
with coupling to external electromagnetic potentials
$A_{\mu}= (A_{i}, A_{0})$
introduced through $\Pi_{i} = -i\partial_{i} + e A_{i}$ 
[$i= (1,2)$ or $(x,y)$]; $v_{\rm F} \sim 10^{6}$ m/s is the Fermi
velocity. Here $\psi$ and $\chi$ stand for the electron fields
near the $K$ and $K'$ points (at wave vectors  
$\mp {\bf q}$) and have the structure 
$\psi_{\alpha}= (\psi_{1}, \psi_{2})^{\rm t} 
\propto (U_{\bf k - q}, V_{\bf k - q})^{\rm t}$ 
and $\chi_{\alpha} = (\chi_{1}, \chi_{2})^{\rm t} 
\propto (-V_{\bf k + q}, U_{\bf k + q})^{\rm t}$,
where $U_{\bf k}=U_{\bf k}(t)$, e.g.,
denotes the Fourier transform of $U({\bf x}, t)$ for short.

For generality we have introduced a tiny $\lq$mass' gap $m$, 
which works to lift the degeneracy of the lowest ($n=0$) Landau level alone.
The possibility of such a mass gap has been
discussed~\cite{NM,AF,Massgap,FL} in connection with the observed
lifting~\cite{ZJS} of the degeneracy of the $n=0,1$ levels in high
magnetic fields. Nonzero mass $m\not=0$ spoils the pseudospin 
(or sublattice) SU(2) symmetry of $H$, which rotates $(\psi,\chi)$. 
Actually, we keep $m\not=0$ to reveal the particle/hole character 
of the  n=0 levels, as remarked below, and set
$m\rightarrow 0$ eventually.

For clarity, we suppress the electron spin, which is treated
as a global SU(2) symmetry of $H$ by doubling the fields,
$\psi^{a}$ and $\chi^{a}$ with $a= (\uparrow, \downarrow)$.
The SU(2)-breaking Zeeman coupling is considerably
weak, compared with the Coulomb interaction, for graphene and is
ignored in the following.

The Coulomb interaction is written as
\begin{equation}
H^{\rm Coul} 
= {1\over{2}} \sum_{\bf p}
v_{\bf p}\, \rho_{\bf -p}\, \rho_{\bf p},
\end{equation}
where $\rho_{\bf p}$ is the Fourier transform of the electron number
density $\rho = \psi^{\dag}\psi + \chi^{\dag}\chi$;
$v_{\bf p}= 2\pi \alpha/(\epsilon_{\rm b} |{\bf p}|)$ is the
Fourier transform of the Coulomb potential 
$v({\bf x})  = \alpha/(\epsilon_{\rm b} |{\bf x}|)$
with the fine-structure constant $\alpha = e^{2}/(4 \pi \epsilon_{0})
\approx 1/137$ and the substrate dielectric constant
$\epsilon_{\rm b}$.  
In this paper we shall focus on the long-wavelength properties of
graphene and consider $H + H^{\rm Coul}$ as our basic Hamiltonian; 
other short-distance corrections~\cite{AF}, 
including the oscillating piece
$\propto e^{\pm 2i {\bf q} \cdot {\bf x}}$ of the Coulomb interaction,
will thus be ignored.

Let us place graphene in a strong magnetic field and 
first study how the electrons in graphene respond 
to weak external potentials $A_{\mu}(x)$. To this end we set 
$A_{i}(x)\rightarrow A^{B}_{i}({\bf x}) + A_{i}({\bf x},t)$ 
in the Hamiltonian~(\ref{Hzero}), where the vector
potential ${\bf A}^{\! B}= B\, (-y,0)$
supplies a uniform magnetic field $B_{z}=B>0$ normal 
to the graphene plane. 
We turn off $H^{\rm Coul}$ for the moment.

When $A_{\mu}=0$, the eigenmodes of $H$ are
Landau levels of $\psi$ and $\chi$ of energy 
\begin{equation}
\epsilon_{n} = s_{n}\,  \omega_{\rm c} 
\sqrt{|n| + \textstyle{1\over{2}} m^{2}\ell^{2}},
\label{energyn}
\end{equation}
labeled by integers $n=0,\pm 1, \pm2, \dots$, and
$p_{x}$ (or $y_{0} \equiv \ell^{2} p_{x}$ with the magnetic length 
$\ell \equiv 1/\sqrt{eB}$); 
$\omega_{\rm c} = \sqrt{2}\, v_{\rm F}/\ell$ is the basic cyclotron
frequency.  Here $s_{n} \equiv {\rm sign}\{n\} (=\pm 1)$ specifies 
the sign of the energy $\epsilon_{n}$.

For each $n\not=0$, $\psi$ and $\chi$ have the same spectra 
and the positive- and negative-energy (i.e., $n>0$ and $n<0$) levels are
symmetric in structure.
[This is a consequence of the conjugation symmetry
${\cal H}_{+} = - \sigma_{3}({\cal H}_{-})\, \sigma_{3}$ 
with $A_{0}=0$.]
The $n=0$ spectra depend on the sign of $m$. 
Let us take $m>0$.
Then the $n=0$ level of $\psi$ has negative energy 
$\epsilon_{0_{-}} = - v_{\rm F} m$ while that of $\chi$ has positive
energy $\epsilon_{0_{+}} = v_{\rm F} m$.  
These $n=0_{\mp}$ levels represent holes and electrons via quantization;
and this hole/particle characterization persists even  in the limit
$m\rightarrow 0$ where the $n=0_{\mp}$ levels become degenerate.  
These $n=0_{\pm}$ eigenmodes have components only 
on each separate sublattice; see Appendix A for details of the wave
functions.   With the electron spin taken into account, each Landau
level is thus four-fold degenerate, except for the 
$n=0_{\pm}$ levels which are doubly-degenerate.

To make this Landau-level structure explicit, it is useful to pass to
the $|n,y_{0}\rangle$ basis, with the expansion 
$\psi ({\bf x}, t) = \sum_{n, y_{0}} \langle {\bf x}| n, y_{0}\rangle\, 
\psi_{n}(y_{0},t)$. (From now on, we shall only display the $\psi$
sector since the $\chi$ sector is 
obtained by reversing the sign of $m$.)
The translation is simple~\cite{GJ,KSproj}: In the $|n, y_{0}\rangle$
representation the coordinate ${\bf x}=(x_{1}, x_{2})$ of 
$A_{\mu}({\bf x},t)$ is split into the center coordinate 
${\bf r} = (r_{1}, r_{2}) = (i\ell^{2}\partial/\partial y_{0}, y_{0})$
with uncertainty 
$[r_{1}, r_{2}] =i\ell^{2}$ and the relative coordinate 
${\bf X} = (X_{1}, X_{2})$ with $[X_{1}, X_{2}] =-i\ell^{2}$ 
which are matrices in the Landau-level index.
The Hamiltonian $H$ thereby is rewritten as
\begin{eqnarray}
H\! &=& \!\! \int\! dy_{0} \!\!\!
\sum_{n,n'=-\infty}^{\infty} \!\!\!
\psi^{\dag}_{n}(y_{0},t)\bigl\{ \epsilon_{n}\,
\delta_{n n'} +e V_{n n'} \bigr\} \psi_{n'}(y_{0},t), \nonumber \\
V_{n n'}\!\! &=& \! v_{\rm F}\,{\cal A}_{n n'}
+  v_{\rm F}\, {\cal A}^{\dag}_{n n'}
-({\cal A}_{0})_{n n'}.
\label{Hzeronn}
\end{eqnarray}
Here the electromagnetic coupling $V_{n n'}$ is a matrix in the level
index and a function of 
${\bf r} = (i\ell^{2}\partial/\partial y_{0}, y_{0})$:
The field 
\begin{equation}
 ({\cal A}_{0})_{n n'}({\bf r}, t) =
\sum_{\bf p} e^{i{\bf p \cdot r}}g_{n n'}({\bf p})e^{-{1\over{4}}\,
\ell^{2} {\bf p}^{2}}\,  (A_{0})_{\bf p}
\label{Azeronn}
\end{equation}
is expressed in terms of the Fourier transform $(A_{0})_{\bf p}(t)$ of 
$A_{0}({\bf x},t)$ and the coefficient matrix
\begin{eqnarray}
g_{n n'}({\bf p}) &=& \textstyle{1\over{2}}\, \Big[
c_{n}^{+}\, c_{n'}^{+}\, 
f_{|n|\! -\!1, |n'| -\!1}({\bf p}) \nonumber\\
&&\ \ \  + s_{n}s_{n'}c_{n}^{-}\, c_{n'}^{-}\, 
f_{|n|, |n'|}({\bf p}) \Big] 
\end{eqnarray}
with $c_{n}^{\pm} = \sqrt{1 \pm  v_{\rm F} m/\epsilon_{n}}$, where
\begin{eqnarray}
f_{k n}({\bf p}) 
&=& \sqrt{{n!\over{k!}}}\,
\Big({i\ell p\over{\sqrt{2}}}\Big)^{k-n}\, L^{(k-n)}_{n}
\Big(\textstyle{1\over{2}} \ell^{2}{\bf p}^{2}\Big)
\end{eqnarray}for $k \ge n$; $ p=p_{y}\! +i p_{x}$;
for $k\le n$ one may set $k\leftrightarrow n$ and 
$p\rightarrow p_{y}\! - ip_{x}$.
Similarly, ${\cal A}_{n n'}$ is defined by Eq.~(\ref{Azeronn}) with
$(A_{0})_{\bf p} \rightarrow  (A_{x} + i A_{y})_{\bf p}$
and 
\begin{equation}
g_{n n'}({\bf p}) \rightarrow j_{n n'}({\bf p}) =
\textstyle{1\over{2}}\,  c_{n}^{+}\, c_{n'}^{-}\,\, 
f_{|n| -1, |n'|}({\bf p});
\label{jnn}
\end{equation}
${\cal A}^{\dag}_{n n'}$ is defined with  $(A_{x} - i A_{y})_{\bf p}$
and ${j}^{\dag}_{n n'}({\bf p}) \equiv j_{n' n}(-{\bf p})$.
It is clear from Eq.~(\ref{Hzeronn}) that $ (A_{0})_{\bf p}$ is coupled
to the charge density
$\rho_{-{\bf p}}(t) =\int d^{2}{\bf x}\,  
e^{i {\bf p\cdot x}}\,\psi^{\dag}
\psi$ rewritten as 
\begin{equation}
\rho_{-{\bf p}} = e^{-{1\over{4}}\, \ell^{2} {\bf p}^{2}}
\sum _{n, n'} g_{n n'}({\bf p})\int dy_{0}\,
\psi_{n}^{\dag}\, e^{i{\bf p\cdot r}}\,
\psi_{n'} ; 
\nonumber
\end{equation}
analogously for the current  $J(x) = {1\over{2}}\,
\psi^{\dag}(\sigma_{1} +i \sigma_{2}) \psi$ coupled 
to $A= A_{x} + i A_{y}$.

Even a weak potential $A_{\mu}(x)$ causes mixing of the Landau levels,
and its effect is calculated by diagonalizing  $H$ 
with respect to the true levels $\{n\}$ by a suitable U($\infty$)
transformation 
$\psi^{G}_{n}(y_{0},t)= \sum_{m} G_{nm}({\bf r},t)\, \psi_{m}(y_{0},t)$.
This yields 
$H =\int dy_{0}\sum_{n}\,
(\psi^{G})^{\dag}_{n} {\cal H}^{G}_{nn} \psi^{G}_{n}$
with
${\cal H}^{G} = G ({\cal H} - i\partial_{t})G^{-1}$ diagonal in the level
index.
To $O(A_{\mu}^{2})$,
${\cal H}^{G}_{nn} \equiv h_{n}= \epsilon_{n} + V_{nn} + V_{n}^{(2)}$
with
\begin{equation}
V_{n}^{(2)}\!
= {1\over2} {\sum}'_{k}\Big\{ V_{nk}
{1\over{\epsilon_{n}\! -\epsilon_{k} +i\partial_{t}}}\, V_{kn} 
- (n \leftrightarrow k) \Big\},
\label{VVresponse}
\end{equation}
where $i\partial_{t}$ acts on $V_{kn}$.
This $O(V^{2})$ term embodies the $\langle \rho\rho \rangle$ 
and $\langle J J^{\dag} \rangle$ correlation functions and 
constitutes the linear response  of the system. 
Actually it only takes account of inter-Landau-level transitions due to
the cyclotron modes; they govern the response of graphene 
at integer filling factors, on which we focus throughout this paper. 
For graphene at noninteger filling one encounters intra-level
(collective) excitations,~\cite{GMP} which are 
less sensitive to long-wavelength probes and are determined by
diagonalizing the Coulomb Hamiltonian $H^{\rm Coul}$ projected  to a
given Landau level.

Let us now consider the long-wavelength part of 
the electromagnetic response, 
which, in view of gauge invariance, takes the form
\begin{eqnarray}
h_{n} &=& \epsilon_{n} - eA_{0} 
+ \gamma_{n}\, {1\over{2}} \epsilon^{\mu\nu\lambda}
A_{\mu}\partial_{\nu}A_{\lambda} \nonumber\\
&&-\beta_{n}\,
{1\over{2}}\, e^{2} (E_{k})^{2} + \eta_{n}\,{1\over{2}} e^{2}
(A_{12})^{2}+ \cdots,
\label{longwavelenghresponse}
\end{eqnarray}
where $E_{k}= -\partial_{k}A_{0} -\partial_{t}A_{x}$ and
$A_{12}=\partial_{1}A_{2} -\partial_{2}A_{1}$; $\epsilon^{012}=1$.
Actual calculations are simplified if we take 
$A_{\mu}({\bf x},t)$ uniform in $x_{1}$, i.e., $A_{\mu}(y,t)$.
Then the CS term
$\epsilon^{\mu\nu\lambda} A_{\mu}\partial_{\nu}A_{\lambda}$
$\sim  (A_{y}\dot{A}_{x} -A_{x}\dot{A}_{y})$,
the $E_{i}^{2} \sim (\dot{A}_{x})^{2} + (\dot{A}_{y})^{2}$ term
and the  $(A_{12})^{2} \sim (\partial_{y}A_{x})^{2}$ term 
are derived from the $O(\partial_{t})$,  $O(\partial_{t}^{2})$ 
and $O({\bf p}^{2})$ parts of the $\langle J J^{\dag} \rangle$ 
response function, respectively.
A direct calculation yields 
\begin{eqnarray}
\gamma_{n} &=& e^{2} \ell^{2},\ \ 
\beta_{n} = \ell^{4}\, {3\epsilon_{n}^{2} - m^{2} v_{\rm F}^{2}
\over{2 v_{\rm F}^{2}\,\epsilon_{n}}}, \nonumber\\
\eta_{n} &=&  \ell^{4}\, (3 + {m^{2}v_{\rm F}^{2}\over{\epsilon_{n}^{2}}})
\,  {(\epsilon_{n}^{2} - m^{2}v_{\rm F}^{2})\over{4\epsilon_{n}}} .
\label{gammabeta}
\end{eqnarray}
Note that $\gamma_{n}$ is independent of
$n$ while $\beta_{-n} = - \beta_{n}$ (with
$\beta_{0_{\mp}}= \mp |m|\ell^{4}/v_{\rm F}$) and
$\eta_{-n} = -\eta_{n}$ (with $\eta_{0}=0$).

We are now ready to introduce the quantum vacuum as the Dirac sea with
all negative-energy electron states occupied.
Let us rename $a_{n}(y_{0},t) = \psi^{G}_{n}(y_{0},t)$ and 
$d_{n}(y_{0},t) = (\psi^{G})^{\dag}_{-n}(y_{0},t)$ with $n>0$ to denote
electrons and holes over the vacuum.
The Hamiltonian then reads
\begin{eqnarray}
H\!\! &=& \!\!\!\int\! dy_{0}
\Big[ \sum _{n=1}^{\infty}\!
h_{n}\, a_{n}^{\dag}a_{n}\! -\!\! \sum _{n=0}^{\infty} h_{-n}
d^{\dag}_{n}d_{n}\! +\triangle \Big],\ \\
\triangle &=& \delta_{0}\,  
\sum _{n=0}^{\infty} h_{-n}, 
\label{vacuumresponse}
\end{eqnarray}
where $\delta_{0} \equiv \delta (y_{0}=0) =
L_{x}/(2\pi \ell^{2})$ stands for the degeneracy of each Landau level;
$L_{x} = \int dx_{1}$.

Here $\triangle$ represents the quantum response of the vacuum.
Apparently it is indeterminate because of the sums over an infinite
number of Landau levels.  Note, e.g., that 
$\gamma_{n} = e^{2} \ell^{2} >0$ for all $n$. 
This would naively mean that the vacuum, the Dirac sea, carries an
infinitely large Hall conductance, which is physically unacceptable. 
To obtain a sensible answer one has to  define the sum carefully.  
For regularization let us truncate the spectrum to
a finite number $(2 N + 1)$ of Landau levels
$\{ n\}$ with $-\epsilon_{N} \le \epsilon_{n} \le \epsilon_{N}$, and
let $N \rightarrow \infty$ at the end.~\cite{fnReg}

It is instructive to see why and how this regularization works physically. 
Let us note that in $V^{(2)}_{n}$ of Eq.~(\ref{VVresponse})
the virtual $(n \rightarrow k \rightarrow n)$ transition and the related
$(k \rightarrow n \rightarrow k)$ transition contribute equally 
but in opposite sign.  For definiteness, we denote  by $F_{n}$ 
the $O(V^{2})$ response, 
$\beta_{n},\gamma_{n}, \eta_{n}, \dots$, collectively, 
and write it as the regularized sum
$F_{n} = \sum_{k=-N}^{N} F_{n}^{k}$ over the contribution 
$F_{n}^{k}$ from  the $n\rightarrow k\rightarrow n$
subprocesses.
Then, in the sum $\sum_{n}F_{n}$ cancellation takes place among 
a majority of terms, owing to the antisymmetry 
\begin{equation}
F_{n}^{k} = - F_{k}^{n},
\label{Fnk}
\end{equation}
which implies pair-wise cancellation between the
$n\rightarrow k\rightarrow n$ and
$k\rightarrow n\rightarrow k$ subprocesses.
Fermi statistics is thus naturally taken care of in the
regularized sum.
Such a pair-wise cancellation of virtual processes is a basic
property of Berry's phase~\cite{Berry,FScs}; and it is also an exact
property~\cite{fnRes} of the density and current response functions
(with also the Coulomb interaction included), as is clear from 
their spectral representations.

The vacuum response is now written as a regularized sum
$\triangle = \delta_{0}\, \sum_{n=0}^{N} h_{-n}$ or
$F^{\rm vac}\equiv\sum_{n>0}^{N} F_{-n}$.
Note here that the regularized sum of $F_{n}$ over all the levels
vanishes,  
$\sum_{n=-N}^{N} F_{n} = \sum_{n}  \sum_{k} F_{n}^{k} =0$.
This fact allows one to cast $F^{\rm vac}$ in another suggestive form
\begin{equation}
\sum_{n=-N}^{n_{\rm f}} F_{n} = {1\over{2}}\,
\Big(\sum_{n=-N}^{n_{\rm f}} F_{n} -
\sum_{n> n_{\rm f}}^{N} F_{n}
\Big).
\label{spectralasymmetry}
\end{equation}
Here we have written the response in a slightly generalized
form: The left-hand side is the response of a many-body state 
with Landau levels occupied up to $n= n_{\rm f}$; the choice 
$n_{\rm f}=0_{-}$ thus yields $F^{\rm vac}$.
This formula expresses the quantum response in terms of an asymmetry
in the spectrum of the occupied and empty levels, weighted with
the response $F_{n}$ per level.

Via regularization the calculation of the response is modified.  
A close look into the matrix elements in Eq.~(\ref{jnn}) shows 
that $\beta_{n}$ and $\gamma_{n}$ come only 
from virtual transitions to the adjacent levels 
($n \rightarrow n\pm 1$) and the related ones across the Dirac sea
$(n\rightarrow  -\{n \pm1\} )$.  As a result,
the expressions for $\gamma_{n}$ and $\beta_{n}$ in Eq.~(\ref{gammabeta})
are valid for $|n|\le N\!-\! 1$ while the boundary contributions
$\gamma_{-N}$ and $\beta_{-N}$ have to be calculated 
from the $-N \rightarrow \pm (N\!-\! 1)\rightarrow -N$ processes. 
For $\gamma_{\mp N}$ the result is
\begin{equation}
\gamma_{\mp N} = -e^{2}\ell^{2}\, (N -
\textstyle{1\over{2}} \pm {1\over{2}}\, m/\epsilon_{N} ),
\end{equation}
which then makes $\triangle^{\gamma}$ finite,
\begin{equation}
\triangle^{\gamma} = \delta_{0}\,  
\sum_{n=0}^{N} \gamma_{-n} ={1\over{2}}\, 
e^{2}\ell^{2}\delta_{0} = L_{x}\, {e^{2}\over{4\pi \hbar}}.
\label{deltagamma}
\end{equation}
This implies that the vacuum would have Hall conductance
equal to half of a filled level, $\sigma^{\rm vac}_{\rm Hall} ={1\over{2}}
e^{2}/h$, and that the Hall effect would arise without real electrons or
holes. This nonzero $\triangle^{\gamma}$ is attributed to a
spectral asymmetry due to the $n=0_{-}$ level if one uses 
Eq.~(\ref{spectralasymmetry}) (since $\gamma_{-N} -\gamma_{N}
\rightarrow 0$), as normally implied by an index theorem~\cite{NS}.
The physical mechanism underlying the vacuum Hall effect was discussed
earlier~\cite{Hal,FScs}.
Actually, the $\chi$ sector contributes to $\triangle^{\gamma}$ in
opposite sign and the vacuum Hall effect does not take place for graphene.

In contrast, the vacuum effect survives for susceptibilities
$\beta_{n}$ and $\eta_{n}$ that change sign with $n$.
For  $\beta_{n}$ the boundary contribution again increases rapidly with $N$,
\begin{equation}
\beta_{\mp N}
\approx  \sqrt{2}\, (\ell^{3}/ v_{\rm F})\, 
\Big[ \pm (N - \textstyle{1\over{2}}+ {1\over{3}}\,\lambda)^{3/2}
+ {1\over{2}} \sqrt{\lambda}\,  \Big], 
\end{equation}
where $\lambda ={1\over{2}}\, m^{2}\ell^{2}$, and makes 
the vacuum electric susceptibility 
$\alpha^{\rm vac}_{\rm e} = (\sum_{n=1}^{N} \beta_{-n}
+\beta_{0_{-}})\, \delta_{0}/L_{x}$ finite,
\begin{eqnarray}
\alpha^{\rm vac}_{\rm e} 
&=& {3\, e^{2}\over{2\pi\, \omega_{\rm c}}}\, 
\Big\{ G(\lambda) -{1\over{3}}\,\sqrt{\lambda}\,  \Big\},
\label{aevac}
\end{eqnarray}
with
\begin{eqnarray}
G(\lambda)
&=&  -\sum _{n=1}^{N-1} \Big\{\sqrt{n +\lambda}
 - {\lambda/3 \over{\sqrt{n +\lambda}}} \Big\}  \nonumber\\
&& + \textstyle{2\over{3}}\,
\Big( N -\textstyle{1\over{2}} + {1\over{3}}\, \lambda \Big)^{3/2}, 
\nonumber\\
&=& -\zeta (-\textstyle{1\over{2}}) 
- \lambda\, {1\over{6}}\, \zeta(\textstyle{1\over{2}}) + \cdots,
\nonumber\\
&\approx& 0.2079 +\lambda \cdot 0.2434 + \cdots.
\end{eqnarray}
Here we have used
\begin{eqnarray}
\sum_{n=1}^{N-1} \sqrt{n}
  - \textstyle{2\over{3}}\,
\big( N -\textstyle{1\over{2}} \big)^{3/2} \!\!
&=& \zeta(-\textstyle{1\over{2}}) \approx -0.2079,  \nonumber\\
\sum_{n=1}^{N-1} 1/\sqrt{n} - 2\sqrt{ N -\textstyle{1\over{2}} }
&=& \zeta(\textstyle{1\over{2}}) \approx -1.4604.
\end{eqnarray}

It is instructive to derive this $\alpha_{\rm e}^{\rm vac}$ 
by selectively summing up the virtual
$(-n \rightarrow  n\pm 1 \rightarrow -n)$ processes from the Dirac sea.
This reproduces $G(0)$ as the sum
\begin{equation}
6\, G(0) = 1+ \sum_{k=1}^{N-1} (\sqrt{n+1}-\sqrt{n})^{3}.
\end{equation}

Formula~(\ref{spectralasymmetry}) also leads to the same
result~(\ref{aevac}). 
Note that this formula admits a further generalization: 
Instead of the symmetric spectral cutoff
one can equally well truncate the positive- and negative-energy spectra 
independently by choosing
$-N\le n \le N'$ so that there are 
$(N+ N' +1)$ Landau levels in total.
Clearly $\beta_{-N}$ and $\gamma_{-N}$ are unchanged 
if one takes $N'\ge N$.

With this generalization in mind let us now examine the magnetic
susceptibility of the vacuum, 
$\alpha^{\rm vac}_{\rm m} = (\sum_{n=1}^{N} \eta_{-n}
+\eta_{0_{-}} )\, \delta_{0}/L_{x}$.
A look into the matrix elements again reveals
that the expression~(\ref{gammabeta}) for $\eta_{n}$ is valid only
for $|n|\le N-2$. Also a direct calculation shows that the virtual
transitions from the bottom of the Dirac sea to positive-energy states
{\em combine}~\cite{fnt} to vanish (for $N\rightarrow \infty$) 
once one takes $N'\ge N+2$, and the response is independent 
of the way one sends $(N, N')\rightarrow \infty$.  
We therefore set $N'\ge N+2$ to calculate the boundary term
$\eta_{-N} +\eta_{-(N-1)}$ and obtain  
\begin{equation}
\alpha^{\rm vac}_{\rm m} 
={3\, e^{2} v^{2}_{\rm F} \over{4\pi\, \omega_{\rm c}}}\,  
G_{\rm m}(\lambda),
\label{amagvac}
\end{equation}
with
\begin{eqnarray}
G_{\rm m}(\lambda)
&=&  -\sum_{n=1}^{N}{n\over{n+\lambda}}\, \Big( \sqrt{n +\lambda}
 +{\lambda/3\over{\sqrt{n +\lambda}}}
\Big)  \nonumber\\
&& + \textstyle{2\over{3}}\,
\Big( N +\textstyle{1\over{2}} - {1\over{3}}\, \lambda \Big)^{3/2} ,
\nonumber\\
&=& -\zeta (-\textstyle{1\over{2}}) 
+ {\lambda\over{6}}\, \zeta({1\over{2}})  + \cdots,  \nonumber\\
&\approx& 0.2079 - \lambda \cdot 0.2434 + \cdots.
\end{eqnarray}

Let us next consider the $A_{0}$ response in $h_{n}$, which probes the
charge of the system.
The observable charge is a difference between the $B\not=0$ and $B=0$
cases.  Such an induced charge is detected by a normal-ordered charge
operator  
$:\psi^{\dag}_{n}\psi_{n}:\,  = {1\over{2}}\, 
[\psi^{\dag}_{n},\psi_{n}]$ and a symmetric spectral cutoff,
with the result~\cite{NS}
\begin{equation}
\triangle Q^{\rm vac} = {e\over{2}}\, {1\over{2\pi}\ell^{2}} =
e^{2}B/(4\pi)
\label{inducedcharge}
\end{equation} 
per unit area.
Alternatively, this induced charge is derived from $\triangle^{\gamma}$ 
in Eq.~(\ref{deltagamma}) if one notes that the variation 
of $\triangle Q$ under the change $B\rightarrow B + A_{12}$ is read
from the CS coupling.
It is thus clear that there is no induced vacuum charge in graphene.

Similarly, one can calculate the observable vacuum energy density
$\epsilon_{\rm vac}=\epsilon_{\rm vac}^{B} -\epsilon_{\rm vac}^{B=0}$ 
from the $\epsilon_{n}$ terms in $h_{n}$ via proper regularization; 
see Appendix B for details.
Here we only remark that
this $\epsilon_{\rm vac}$ provides another way to derive 
$\alpha^{\rm vac}_{\rm m}$: Setting $B \rightarrow B + A_{12}$ in
$\epsilon_{\rm vac}$ and expanding it to second order in
$A_{12}$ yields precisely the regularized expression~(\ref{amagvac}) 
for $\alpha^{\rm vac}_{\rm m}$.

We are now ready to write down the electromagnetic response of graphene.  
Let us suppose that the electrons fill up an integral number $\nu$
of Landau levels, with uniform density
$\langle \rho\rangle \equiv \bar{\rho} = \nu /(2\pi \ell^{2})$;
the case of holes is treated analogously.
We take both spin and pseudospin into account, 
and write $\nu = \sum_{n} \nu_{n}$ in terms of the filling factors
$\nu_{n}$ of the $n$th level 
$[\, 0\le \nu_{n} \le 4$ for $n\ge 1$ and 
$0\le \nu_{0_{+}} \le 2\, ]$.
The long-wavelength response of graphene is then summarized
by the Lagrangian
\begin{eqnarray}
L_{A} &=&  \bar{\rho}\, e A_{0}
- e^{2}\ell^{2}\bar{\rho}\, {1\over{2}}\,
\epsilon^{\mu\nu\rho}\, A_{\mu}\partial_{\nu}A_{\rho}
\nonumber\\
&& + {1\over{2}}\,
\alpha_{\rm e}(\nu)\,  {\bf E}^{2} -{1\over{2}}\,\alpha_{\rm
m}(\nu)\, (A_{12})^{2}
\label{emresponse}
\end{eqnarray}
with the susceptibilities
\begin{eqnarray}
\alpha_{\rm e}(\nu)
&=&  {3 e^{2}\over{2\pi \omega_{\rm c} }}\,
\Big[\sum_{n=1}^{\infty} 
\nu_{n} \sqrt{n} + 4\, G(0) \Big] ,
\label{alphaenu} \\
\alpha_{\rm m}(\nu) &=& (v_{\rm F}^{2}/2)\, 
\alpha_{\rm e}(\nu).
\end{eqnarray}
For clarity we shall from now on display formulas with $m\rightarrow 0$.
The vacuum susceptibility $\alpha^{\rm vac}_{\rm e}$ is
almost comparable to the contribution of a single
filled $n=1$ level, i.e., $4 \times 0.21$ vs 1.
Curiously, in contrast, the $n=0_{\pm}$ levels hardly contribute to 
$\alpha_{\rm e}(\nu)$;
their tiny fraction $\propto (\nu_{0_{+}}-2) (\sqrt{2} /3)\, |m| \ell$ 
in $\alpha_{\rm e}$ shows that a mass gap works to reduce 
$\alpha_{\rm e}$, as is clear intuitively.

\section{Long-wavelength effective theory}

In this section we derive an effective gauge theory of graphene
in a magnetic field. 
It is constructed, via functional bosonization~\cite{funcBos}, so as to
reproduce the response~(\ref{emresponse}).  Such an
effective theory has some advantages~\cite{funcBos,LZ}: It allows one to
handle the Coulomb interaction exactly.
It readily admits inclusion of vortices, 
and is thus applicable to the description of the fractional QH effect as
well.  It furthermore unifies~\cite{KSbos} the composite-boson~\cite{CB}
and composite-fermion~\cite{CF} descriptions of the fractional QH effect.

Applying~\cite{KSbos} the standard procedure of functional
bosonization to the response~(\ref{emresponse})  yields an effective
theory of a vector field $b_{\mu}=(b_{0}, b_{1}, b_{2})$, 
with the Lagrangian to $O(\partial^{2})$,
\begin{eqnarray}
L_{b} 
&=& {1\over{2\ell^{2}\bar{\rho}}}\, b_{\mu}\, \epsilon^{\mu\nu\lambda}
\partial_{\nu} b_{\lambda} + {1\over{\ell^{2}}}\, b_{0} \nonumber\\
&&
+ {1\over{2\ell^{2}\bar{\rho}\, \omega_{\rm eff}}}\, 
\Big\{ (b_{k0})^{2} - {1\over{2}}\, v_{\rm F}^{2}\, (b_{12})^{2} \Big\} 
\end{eqnarray}
and the {\sl effective} cyclotron frequency 
\begin{eqnarray}
\omega_{\rm eff} &=& e^{2}\ell^{2}\bar{\rho} /\alpha_{\rm e}(\nu)
=\omega_{\rm c} g(\nu),  \nonumber\\
g(\nu) &\approx& {1\over{3}}\, 
{\nu \over{ \sum_{n=0}^{\infty} \nu_{n} \sqrt{n} + 4\, G(0)} } ,
\label{weff}
\end{eqnarray}
where $b_{\mu\nu} = \partial_{\mu}b_{\nu} -
\partial_{\nu}b_{\mu}$;
$\ell^{2}\bar{\rho} =  \nu/(2\pi)$ and $\nu= \sum_{n=0}^{\infty}
\nu_{n}$.
The Coulomb interaction and coupling to $A_{\mu}$ are incorporated into
this theory, with the Lagrangian
\begin{equation}
L_{\rm eff}[b] = -e A_{\mu}\, \epsilon^{\mu\nu\lambda}
\partial_{\nu} b_{\lambda}
- {1\over{2}}\, \delta b_{12}\, v \, \delta b_{12}+ L_{b},
\label{Leffb}
\end{equation}
where $\delta b_{12}\, v \, \delta b_{12} = \int d^{2}{\bf x}'
\delta b_{12}(x)\, v({\bf x\! -\! x'}) \delta b_{12}(x')$ for short and
$\delta b_{12} = b_{12} - \bar{\rho}$.

This effective theory reproduces the original response~(\ref{emresponse}),
and actually more:  The $A\epsilon\partial A$, ${\bf E}^{2}$ and
$(A_{12})^{2}$ terms acquire a common kernel
so that, e.g.,  
${\bf E}^{2} \rightarrow  {\bf E}\, {\cal K}\, {\bf E}$, 
with
\begin{equation}
{\cal K} = \omega^{2}_{\rm eff}/[\omega^{2}_{\rm eff} +
\partial_{t}^{2} - (\omega_{\rm eff} \ell^{2} \bar{\rho}\,v_{\bf p}
+\alpha_{\rm m}/\alpha_{\rm e})\,
\nabla^{2} ].
\end{equation}
This shows that the Coulomb interaction $v_{\bf p} \sim
e^{2}/|{\bf p}|$ substantially modifies the dispersion of 
the cyclotron mode at long wavelengths ${\bf p} \rightarrow 0$,
\begin{equation}
\omega ({\bf p})
\approx  \omega_{\rm eff} 
+ {1\over{2}}\, (\ell^{2}\bar{\rho}\, v_{\bf p} 
+ {v_{\rm F}^{2}\over{2\omega_{\rm eff}}} )\, {\bf p}^{2}.
\label{Coulombshift}
\end{equation}

For graphene the Landau levels are not equally spaced and the excitation
gaps depend on the level index $n$ or $\nu$.
When the $n$th levels are occupied, 
i.e., at $\nu= 4n +2$, the minimum activation gap is
$\triangle\omega_{\rm c}^{(n)}= \epsilon_{n+1} - \epsilon_{n}
= (\sqrt{n+1}  -\sqrt{n} )\, \omega_{\rm c}$;
numerically, $\triangle\omega_{\rm c}^{(0)} = 1.0\,  \omega_{\rm c}$,
$\triangle\omega_{\rm c}^{(1)} \approx 0.4142\, \omega_{\rm c}$,
$\triangle\omega_{\rm c}^{(2)} \approx 0.3178\, \omega_{\rm c}$,
$\triangle\omega_{\rm c}^{(3)} \approx 0.2679\, \omega_{\rm c}$,
for $\nu= 2, 6, 10, 14$, respectively.
The $\omega_{\rm eff}$ in Eq.~(\ref{weff}) represents such an excitation
gap.

For $\nu=2$, $g(2) =1/\{ 6 G(0)\} \approx 0.802$, which comes solely from
the vacuum fluctuations, appreciably deviates from the number 1.0 quoted
above. For $\nu=6$, $g(6) \approx 0.4141$;
for $\nu=10, 14, \dots$, $ g(\nu)$ almost precisely agrees
with the quoted numbers, and the agreement becomes exact for
$\nu\rightarrow \infty$.
It is somewhat surprising that an effective theory constructed from
the long-wavelength response alone gives an excellent description of 
the excitation spectrum.

In other words, the effective theory reveals a general link: 
The susceptibility $\alpha_{\rm e}$ of a QH system is essentially
determined  by a Hall conductance and the Landau gap,
\begin{equation}
\alpha_{\rm e}(\nu) = (e^{2}/2\pi)\, 
\nu/\omega_{\rm eff} = \sigma_{\rm Hall}/
\triangle\omega_{\rm c}^{(n)}.
\label{aevsSigmaHall}
\end{equation}
This implies that $\alpha_{\rm e}(\nu)$  rises linearly with $\nu$ 
for typical QH systems with equally-spaced Landau
levels while it grows more rapidly for graphene, 
\begin{equation}
\alpha_{\rm e}(\nu) \propto \nu^{3/2}.
\end{equation}
Actually, formula~(\ref{aevsSigmaHall})
is a condensed-matter realization of 
the related fact~\cite{massgeneration} in relativistic field theory
that in 2+1 dimensions the gauge field acquires a mass from 
the CS term in the presence of  the  Maxwell term $\sim {\bf E}^{2}$.

\section{Polarization function}

We have seen that the vacuum component $\alpha_{\rm e}^{\rm vac}$
constitutes an appreciable portion of the susceptibility 
$\alpha_{\rm e}(\nu)$. In this section we examine the effect of
polarization at shorter wavelengths and extract some
observable consequences.  
We shall see that the screening properties of graphene in a magnetic
field are substantially different from those of graphene without the
magnetic field, discussed recently~\cite{Ando}.

Let us turn off the Coulomb interaction for the moment and calculate 
the polarization function
$P({\bf p}, \omega) \equiv  -i\, \langle \rho (x) \rho (y)
\rangle^{\rm F.T.}$,
where F.T. stands for the Fourier transform.
To this end we denote by 
$P_{n}({\bf p}, \omega) 
= -i\, \langle n|T \rho (x) \rho (y) |n \rangle^{\rm F.T.}$ 
the polarization function for the filled
$n$th Landau level $|n\rangle = \{ |n, y_{0}\rangle\}$ and 
write it as a regularized sum 
$P_{n}({\bf p}, \omega) = \sum_{k=-N}^{N} P_{n}^{k}({\bf p}, \omega)$
over components $P_{n}^{k}$ coming from the virtual 
($n\rightarrow k\rightarrow n$) transitions.  
As before, we truncate the spectrum to a finite interval 
$-N\le n, k \le N$.
From Eq.~(\ref{VVresponse}) one finds 
\begin{eqnarray}
P^{k}_{n}({\bf p}, \omega)
&=& -\Big\{ {1\over{\epsilon_{kn} -\omega}}
+ {1\over{\epsilon_{kn} +\omega}} \Big\}\, \sigma_{n}^{k}({\bf p}),
\\ 
\sigma_{n}^{k}({\bf p})&=& {1\over{2\pi \ell^{2}}}\,
e^{- {1\over{2}}\,\ell^{2}{\bf p}^{2}}g_{nk}({\bf p})\, g_{kn}(-{\bf p}),
\end{eqnarray}
where $\epsilon_{kn} = \epsilon_{k} - \epsilon_{n}$;
$\sigma_{n}^{k}({\bf p})$ is the spectral weight.

One can calculate the response 
$P({\bf p}, \omega) = \sum_{n}P_{n}({\bf p}, \omega)$ 
by selectively summing $P^{k}_{n}$ over occupied levels $\{n\}$ and
unoccupied levels $\{k\}$ for a given graphene state.
In practice, it is advantageous, especially for numerical evaluations, 
to calculate $P_{n} = \sum_{k=-N}^{N} P_{n}^{k}$ by
summing over all levels $\{k\}$ 
since Fermi statistics is automatically taken care of in the
regularized sum, as noted in the previous section.  
We focus on the real part of $P_{n}({\bf p}, \omega)$ in the static limit
$\omega\rightarrow 0$ below.  
For the $\psi$ sector the result is
\begin{eqnarray}
P_{n}({\bf p}, 0) 
&=&  - {3\,{\bf p}^{2}\over{2\pi\, \omega_{\rm c} }}\,
 s_{n}\, \hat{\xi}_{n}({\bf p}^{2}),\\
\hat{\xi}_{n}({\bf p}^{2}) &=&  e^{-{1\over{2}}\,
\ell^{2}{\bf p}^{2}} \sqrt{|n|}\, 
\xi(\textstyle{1\over{2}}\, \ell^{2}{\bf p}^{2},|n|), 
\label{Pnpzero}
\end{eqnarray}
with  
\begin{eqnarray}
&&\!\!\!\!\!\!\!
6\, \xi(x,n)  \nonumber\\
&=& \! \! \sum_{k=1}^{N-n}  
 {(n -\! 1)! \over{(n \!+\! k \!-\! 1)!}}\, { x^{k-1}\over{k}}
\Big[  (L^{(k)}_{n}\! +\! L^{(k)}_{n-1})^{2} \!
-{k\over{n\!+\! k}} (L^{(k)}_{n})^{2} \Big]\
\nonumber\\
&&\! - \sum_{k=1}^{n-1}  
 {(n -\! k)! \over{n!}}\, { x^{k-1}\over{k}}\,
\Big[  (L^{(k)}_{n -k} +L^{(k)}_{n-k-1})^{2} \nonumber\\
&&\ \ \ \ \ \ \
+{k\over{n\!-\! k}}\, (L^{(k)}_{n-k-1})^{2} \Big]
\nonumber\\ 
&&
- {x^{n-1} \over{n!\, n}}
-{1\over{4 n x}}\, (L_{n} -L_{n-1})^{2} 
\ \ \ \  ({\rm for}\ n \ge 1),
\label{xin} 
\end{eqnarray}
and $\xi(x,n=0)=0$, where $L^{(k)}_{n}=L^{(k)}_{n}(x)$ for short;
$\xi(x,n)$ are normalized so that $\xi(0,n)=1$ for $1\le n\le N-1$, with
their reference to the cutoff $N$ suppressed. Note that $P_{-n}= -P_{n}$;
in particular, $P_{0}=0$, i.e., the $n=0_{\pm}$ levels do not contribute
to polarization  (for $m \rightarrow 0$).

One can now write $P({\bf p},0)$ or the static susceptibility function
$\alpha_{\rm e}[{\bf p}] = - (e^{2}/{\bf p}^{2})\, P({\bf p},0)$
for graphene as
\begin{eqnarray}
\alpha_{\rm e}[{\bf p}] 
&=&  {3\,e^{2}\over{2\pi\, \omega_{\rm c} }}\,
\, \Big[ \sum_{n>0} \nu_{n}\, \hat{\xi}_{n}({\bf p}^{2}) 
+ 4\, \hat{\xi}^{\rm vac}({\bf p}^{2}) \Big],
\end{eqnarray}
where $ \hat{\xi}^{\rm vac}({\bf p}^{2})=
- \sum_{n=1}^{N} \hat{\xi}_{n}({\bf p}^{2})$.
This $\alpha_{\rm e}[{\bf p}]$ is correctly reduced to 
$\alpha_{\rm e}$ in Eq.~(\ref{alphaenu}) 
for ${\bf p} \rightarrow 0$.

Let us now turn on the Coulomb interaction 
$v_{\bf p} = 2\pi \alpha/(\epsilon_{\rm b} |{\bf p}|)$ and 
study its effects in the RPA.  
The RPA dielectric function is written as 
\begin{equation}
\epsilon ({\bf p}, \omega) 
= 1 - v_{\bf p}\, P({\bf p}, \omega), 
\end{equation}
and the RPA polarization function
$P_{\rm RPA}({\bf p},\omega) =
P({\bf p},\omega)/\epsilon ({\bf p},\omega)$ 
is screened through $\epsilon ({\bf p},\omega)$.
The static dielectric function then reads
\begin{equation}
\epsilon ({\bf p},0)
= 1 + {3\alpha \over{\epsilon_{\rm b}\, v_{\rm F}}}\,  
{\ell |{\bf p}|\over{\sqrt{2}}}\,
\Big[ \sum_{n>0} \nu_{n}\, \hat{\xi}_{n}({\bf p}^{2}) 
+ 4\, \hat{\xi}^{\rm vac}({\bf p}^{2}) \Big], 
\nonumber
\end{equation}
where $\alpha = e^{2}/(4\pi \epsilon_{0}) \approx 1/137$
and $v_{\rm F} \approx c/300$. 
With air on one side of the graphene plane and SiO$_{2}$ on the other,
the unscreened dielectric constant is estimated~\cite{AF} to be
$\epsilon_{\rm b} \approx 2/(\epsilon_{\rm air}^{-1} 
+ \epsilon_{{\rm SiO}_{2}}^{-1}) \approx  1.6$ 
with $\epsilon_{\rm air}\approx 1$ and
$\epsilon_{{\rm SiO}_{2}}\approx 4$;
this yields an estimate of the coefficient 
\begin{equation}
3\alpha/ (\epsilon_{\rm b}\, v_{\rm F}) 
=3\sqrt{2}\, \alpha/(\epsilon_{\rm b}\ell \omega_{\rm c})
\approx 4.1\ .
\label{coeff}
\end{equation}


\begin{figure}[tbp]
~\vskip1cm
\begin{center}
\scalebox{0.9}{
\includegraphics{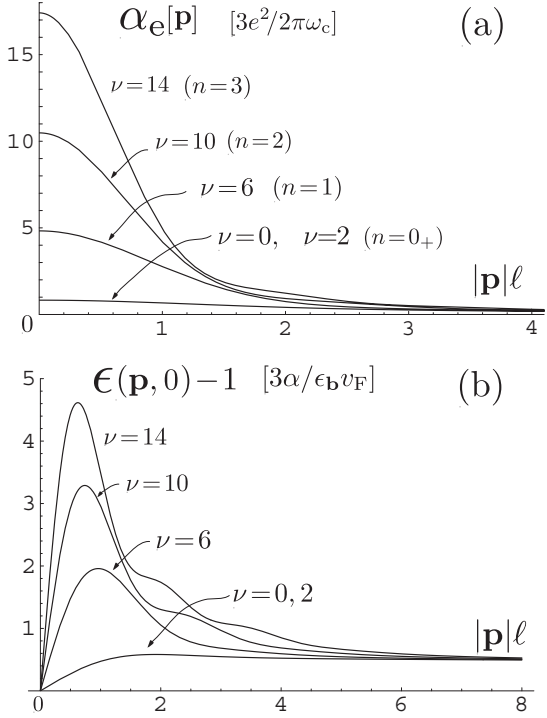}}
\end{center}
\caption{ (a) Susceptibility $\alpha_{e}[{\bf p}]$ for graphene 
at integer fillings $\nu$, plotted in units of 
$3e^{2}/(2\pi \omega_{\rm c})$;
$n$ refers to the highest occupied Landau level.  
(b) Dielectric function $\epsilon({\bf p},0) -1$ is plotted 
in units of $3\alpha/ (\epsilon_{\rm b}\, v_{\rm F}) \approx 4.1$ 
so that the peak value of $\epsilon({\bf p},0)$ ranges from 3.4 to 20 
for $\nu=0 \sim 14$.}
\end{figure}


\begin{figure}[htbp]
\begin{center}
\scalebox{1}{
\includegraphics{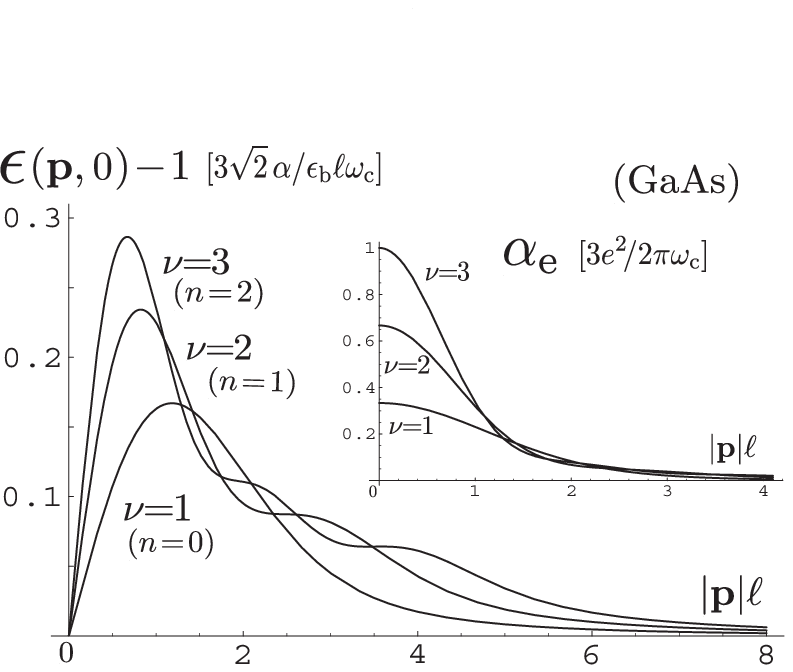}}
\end{center}
\caption{
Susceptibility $\alpha_{e}[{\bf p}]$ and 
$\epsilon({\bf p},0) -1$ for a GaAs system (per spin).
The peak value of $\epsilon({\bf p},0)$ ranges from $1.7 \sim 2.2$
for $\nu = 1 \sim 3$ if one takes
$3\sqrt{2}\, \alpha/(\epsilon_{\rm b}\ell \omega_{\rm c}) \approx 4$.
}
\end{figure}


In Fig.~1 we plot $\alpha_{\rm e}[{\bf p}]$ and 
$\epsilon({\bf p},0) -1$ for $\nu=$ 0, 2, 6, 10, 14.
Note first that there is no screening at long distances, 
$\epsilon({\bf p},0)  \rightarrow 1$ for ${\bf p}\rightarrow 0$, 
as is typical of two-dimensional systems.
However, 
$\epsilon({\bf p},0)$ grows rapidly with $|{\bf p}|$ and becomes sizable
for $|{\bf p}|\ell \sim 1$, with its peak value $\epsilon({\bf
p},0)\approx 3.4
\sim 20$ for $\nu = 0 \sim 14$. 
For comparison, analogous plots for a standard (GaAs) QH system~\cite{KH}
(per spin) are shown in  Fig.~2. Both $\alpha_{\rm e}[{\bf p}]$ and
$\epsilon({\bf p},0) -1$ look qualitatively similar in the two cases, 
but there are clear differences: First, for graphene even the vacuum 
(with $\nu=0$) has a finite polarization,
which decreases only gradually with increasing $|{\bf p}|\ell$;
this reflects the fact that the Dirac sea has quantum
fluctuations of a whole range of wavelengths.
Second, for graphene $\alpha_{\rm e}[{\bf p}]$
and $\epsilon({\bf p},0)$ grow fast with $\nu$. 
This reflects the Landau-level characteristic of graphene;
the quantum effect grows as the level gaps become narrower.

To compare graphene and GaAs systems let us note the following:
For GaAs $\omega_{\rm c} = eB/m^{*} \approx 20\, B$[T] K  
is considerably smaller than for graphene
while the substrate dielectric constant 
$\epsilon_{\rm b} \approx 12.9 \gg  (\epsilon_{\rm b})_{\rm graphene} 
\approx 1.6$
so that
\begin{equation}
(\epsilon_{\rm b}\, \omega_{\rm c})_{\rm GaAs}/
(\epsilon_{\rm b}\, \omega_{\rm c})_{\rm graphene} \approx 0.4\, 
\sqrt{B[{\rm T}]}.
\label{GaAsvsgraph}
\end{equation}
Figure~2 indicates that the effect of polarization is generally not quite
appreciable for the GaAs system, with 
$\epsilon({\bf p},0)\approx 1.7 \sim 2.2$ at $|{\bf p}|\ell \sim 1$ for
$\nu = 1 \sim 3$ if we take
 $3\sqrt{2}\, \alpha/(\epsilon_{\rm b}\ell \omega_{\rm c}) \approx 4$,
which, in view of Eqs.~(\ref{coeff}) and (\ref{GaAsvsgraph}), would
overestimate $\epsilon({\bf p},0)$ numerically  
over a typical range of the applied magnetic field $B>10$ T.

One can derive~\cite{KH} the excitation spectrum corrected by the Coulomb
interaction from the RPA response 
$P_{\rm RPA}({\bf p},\omega) =P({\bf p},\omega)/\epsilon ({\bf p},
\omega)$.  Let us isolate from $P({\bf p}, \omega)$ one of its poles at
$\omega \sim \epsilon_{kn} = \epsilon_{k} -\epsilon_{n}$ 
and set  $\epsilon ({\bf p}, \omega)  = 1 - v_{\bf p}\, P({\bf p},
\omega) \rightarrow 0$.
This fixes the pole position of the RPA response function, 
$\epsilon_{k,n}^{\rm RPA} = \epsilon_{k} -\epsilon_{n} + \triangle
\epsilon_{k,n}({\bf p})$ with 
\begin{eqnarray}
\triangle \epsilon_{k,n}({\bf p})
&\approx& {\alpha\over{\epsilon_{\rm b}\ell}}\, 
\nu_{\rm g}({\bf p})\, \ell |{\bf p}|/\{2\, \epsilon ({\bf p},0) \}, \\
\nu_{\rm g}({\bf p})\! &=&\!\!\! \sum g_{nk}(-{\bf p})
g_{kn}({\bf p})\, e^{-x}/x,
\end{eqnarray}
where $x= {1\over{2}}\, \ell^{2}{\bf p}^{2}$; the sum is taken over
possible $(k,n)$ that lead to the gap $\epsilon_{k} -\epsilon_{n}$.
We have parametrized $\triangle \epsilon_{k,n}({\bf p})$
so that $\nu_{\rm g}(0)$ corresponds to $\nu = 2\pi\, \ell^{2} \bar{\rho}$
in the excitation spectrum~(\ref{Coulombshift}) of the effective theory.
Note that $\triangle \epsilon_{k,n}({\bf p}\rightarrow 0) = 0$ so that 
the excitation gap $\epsilon_{k} -\epsilon_{n}$ remains unshifted for 
${\bf p}\rightarrow 0$.  
The excitation gap
$\triangle\omega_{\rm c}^{(0)}=
\omega_{\rm c}$, in particular, is realized via the
$0_{\pm} \rightarrow 1$ or $-1 \rightarrow 0_{+}$ transitions,
yielding 
\begin{equation}
\nu_{\rm g}({\bf p}) = 2\, e^{-{1\over{2}}\,\ell^{2}
{\bf p}^{2}}\ \ {\rm for}\ \ \nu=0, 2\ .
\label{nugp}
\end{equation}  
The next gap $\triangle\omega_{\rm c}^{(1)} \approx 0.14\, \omega_{\rm c}$
at $\nu=6$ is due the $1 \rightarrow 2$ transition,
giving 
$\nu_{\rm g}({\bf p}) =  (1 + \sqrt{2}-  x/\sqrt{2} )^{2}\,
e^{-x}$ with $x= {1\over{2}}\,\ell^{2}{\bf p}^{2}$.
Also the gaps $\triangle\omega_{\rm c}^{(n)} 
= (\sqrt{n+1} -\sqrt{n})\, \omega_{\rm c}$ open up 
at $\nu=4n+2$ via the $n\rightarrow n+1$ transitions with  
$\nu_{\rm g}(0) = (\sqrt{n} +\sqrt{n+1})^{2}$.
Actually, these numbers $\nu_{\rm g}(0)$ and $\nu=4n+2$ agree within 3
percent for $n=1$ and the agreement improves rapidly for larger $n$.
This shows that the excitation spectrum~(\ref{Coulombshift}) 
in the effective theory is practically in good agreement with the
RPA result.


\begin{figure}[htbp]
\begin{center}
\scalebox{1}{
\includegraphics{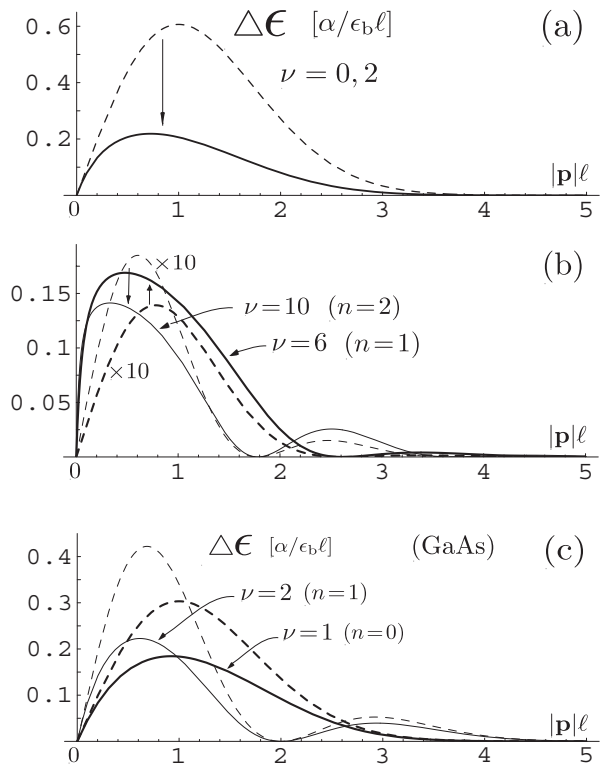}}
\end{center}
\caption{(a) Excitation spectrum near the gap $\omega_{\rm c}$
at $\nu= 0$ and $2$,
plotted in units of $\alpha/\epsilon_{\rm b}\ell$.
It is reduced from the bare spectrum (dashed curve) via screening.
(b) Excitation spectra near the gap $0.41\, \omega_{\rm c}$ and
$0.32\, \omega_{\rm c}$ at $\nu=6$ and $\nu=10$, respectively.
The unscreened spectra (dashed curves) are one order of magnitude larger
than those depicted in the figure. 
(c)  Excitation spectra for a GaAs system per
spin, with the choice $3\sqrt{2}\, \alpha/(\epsilon_{\rm b}\ell
\omega_{\rm c}) \approx 4$.
}
\end{figure}


In Fig.~3 (a) and (b) we plot the exciton spectra $\triangle
\epsilon_{k,n}({\bf p})$ near the energy gaps $\omega_{\rm c}$, 
$0.41\, \omega_{\rm c}$ and $0.32\, \omega_{\rm c}$.
The unscreened spectra (dashed curves) are generally peaked 
around $|{\bf p}| \ell \sim 1$ and are efficiently reduced 
to the screened spectra (solid curves) through 
$\epsilon ({\bf p},0)$ also peaked at $|{\bf p}| \ell \sim 1$.
Near the gap $\omega_{\rm c}$ at $\nu= 0, 2$  
the spectrum is about one third of the unscreened one in
magnitude.  The effect of screening is far more prominent 
at higher $\nu$.
Near the gap $0.41\, \omega_{\rm c}$ at $\nu=6$ the spectrum 
is about one order of magnitude smaller than the unscreened one.  
Near the gap $0.32\, \omega_{\rm c}$ 
at $\nu=10$ the spectrum is even lower than the $\nu=6$
spectrum although the unscreened spectra are opposite in magnitude.

For comparison Fig.~3 (c) shows the exciton spectra in the GaAs
QH system~\cite{KH} near the energy gap $\omega_{\rm c}= eB/m^{*}$ 
at $\nu=1, 2$.  There the spectra are reduced roughly by 50 percent or
less. Clearly the effect of screening is not quite important for GaAs
systems.

\section{Summary and discussion}

In this paper we have examined the electromagnetic characteristics of
graphene in a magnetic field to explore possible signatures of quantum
field theory in the low-energy dynamics of graphene. 
The key features that distinguish graphene from 
standard QH systems are 
(1) the quantum nature of the graphene ground state (vacuum) 
and (2) the relativistic pattern of Landau levels 
$(\propto \pm \sqrt{|n|})$ with gaps decreasing with the level index $n$. 

The graphene vacuum is a dielectric medium
with electron and hole pairs created (from the depth of the Dirac sea) 
in response to an applied field.
In particular, the electric and magnetic susceptibilities 
$(\alpha_{\rm e}, \alpha_{\rm m})$ and the dielectric function
$\epsilon({\bf p}, \omega)-1$ are nonzero for the vacuum 
over a whole range of wavelengths and grow prominently
with increasing filling factor $\nu$ in a way that reflects 
the Landau-level characteristics (2) of graphene.

As for experimental verification, detection of the inter-Landau-level
cyclotron-mode excitations, or excitons,  from the $\nu=0$ ground state
through light absorption or reflection would be a direct signal for
the quantum nature of the vacuum.
A detailed study of the exciton spectra from the states 
with $\nu=0, 2, 6,\dots$ would also reveal the effect of screening 
around $|{\bf p}| \ell \sim 1$,
which is expected to be sizable for graphene, as discussed in Sec.~IV.

 In Sec.~II special emphasis has been placed on the need for
regularization in calculating quantum corrections for graphene.
We have pointed out that the regularized (linear) response
is written as an asymmetry in the spectra of the
occupied and unoccupied states, weighted with the response per state; 
see Eq.~(\ref{spectralasymmetry}).
For the induced CS coupling this spectral asymmetry is related to 
the $\eta$-invariant of the Dirac Hamiltonian (i.e., the difference 
in number between positive and negative eigenvalues~\cite{NS}), although it
vanishes for graphene eventually.   For the susceptibilities or the
polarization function, in contrast, there is no such underlying index
theorem but the spectral asymmetry yields nonzero results for them,  with
the contribution from each level adding up. In this connection, some
peculiarities of the
$n=0_{\pm}$ Landau levels would be worth noting. While they carry the
(normal) Hall conductance 
$\pm e^{2}/h$ per level, they hardly contribute to
susceptibilities $\alpha_{\rm e}$ and $\alpha_{\rm m}$ 
(for $m\rightarrow 0$).   As a result, it is essentially the vacuum state
that would govern  the dielectric property of graphene for $|\nu| \le 2$;
the effect of polarization would be insensitive to carrier densities over
this range; see, e.g., Eq,~(\ref{nugp}).

In Sec.~III we have seen that an effective gauge theory of graphene,
constructed from the long-wavelength response, gives an excellent
description of the excitation spectrum.
In this effective theory graphene and other QH systems look
quite similar, except that the effective Landau gap 
$\triangle\omega_{\rm c} \sim \omega_{\rm eff}$  in Eq.~(\ref{weff}) 
now depends on $\nu$ for graphene. 
We have thereby noted a general relation
$\alpha_{\rm e}  \approx \sigma_{\rm Hall}/ \omega_{\rm eff}$ which
relates the susceptibility $\alpha_{\rm e}$ to the Hall conductance and
the Landau gap.

\acknowledgments

The author wishes to thank A Sawada for useful discussion. 
This work was supported in part by a Grant-in-Aid for Scientific Research
from the Ministry of Education Science and Culture of Japan (Grant No.
17540253).

\appendix

\section{Free Dirac electrons}

In this appendix we summarize the eigenmodes of the Dirac
Hamiltonian in a magnetic field. 
Let us focus on the $\psi$ field governed by 
${\cal H}_{+}= v_{\rm F}\,  (\sigma_{1}\Pi_{1} 
+ \sigma_{2}\Pi_{2} + m \sigma_{3})$ in Eq.~(\ref{Hzero}) with 
$A_{i} \rightarrow {\bf A}^{\! B}=  B\, (-y,0)$. 
The spectrum of $\psi$ consists of an infinite tower of Landau levels
of energy $\epsilon_{n} = s_{n}\,  \omega_{\rm c} \sqrt{|n| + 
m^{2}\ell^{2}/2}$ labeled by integers
$n=0, \pm 1.\pm2, \cdots$, 
with
$\ell=1/\sqrt{eB}$ and $s_{n} \equiv {\rm sign}\{n\} =\pm 1$.
The positive-energy ($n>0$) and negative-energy ($n<0$) eigenmodes 
are neatly cast in the unified form 
\begin{equation}
\psi_{n y_{0}}({\bf x})= {1\over{\sqrt{2}}}
\left(
\matrix{
c_{n}^{+}\, \phi_{|n|-1}(y-y_{0}) \cr
 s_{n} c_{n}^{-}\, \phi_{|n|}(y-y_{0}) \cr}  \right)
{1\over{\sqrt{2\pi \ell^{2}}}}\, e^{ixy_{0}/\ell^{2}},
\label{psinyzero} 
\end{equation}
where $y_{0}\equiv \ell^{2} p_{x}$ and
$c_{n}^{\pm} = \sqrt{1 \pm v_{\rm F} m/\epsilon_{n}}$, in
terms of the standard wave functions 
$\propto \phi_{n}(y- y_{0})\, e^{ixp_{x}}$ of the nonrelativistic
case.
For $m>0$ the $n=0$ eigenmodes
\begin{equation}
\psi_{0_{-}, y_{0}}({\bf x}) = 
\left(
\matrix{
0 \cr -\phi_{0}(y\!-\! y_{0}) \cr}  \right)
{1\over{\sqrt{2\pi \ell^{2}}}}\, e^{ixy_{0}/\ell^{2}}
\end{equation}
 describe negative-energy electrons 
with $\epsilon_{0_{-}} = - v_{\rm F}\, m$. 
For the parity partner $\chi$, 
the spectrum and eigenmodes are obtained by reversing the sign of $m$. 
The $n=0$ modes are actually the $n=0_{+}$ modes  
$\chi_{0_{+}, y_{0}}({\bf x}) \sim (0, \phi_{0})^{\rm t}$ which describe
electrons with positive energy 
$\epsilon_{0_{+}} = v_{\rm F}\, m$.

\section{vacuum energy}

In this appendix we outline the calculation of the vacuum energy 
in a magnetic field $B$, discussed somewhere below 
Eq.~(\ref{inducedcharge}).
Via normal ordering in $H$,
one obtains the vacuum energy per unit area
\begin{eqnarray}
\epsilon_{\rm vac}^{B}= {1\over{2\pi \ell^{2}}}\,(-{1\over{2}})
( 2\sum_{n=1}^{N} \epsilon_{n} + |m|v_{\rm F}).
\label{vacuumE}
\end{eqnarray}
This is to be compared with the energy density in the ordinary  
$(B=0)$ vacuum,
$\epsilon_{\rm vac}^{B=0} = (-{1\over{2}})2 v_{\rm F} \sum_{\bf k} 
\sqrt{{\bf k}^{2} + m^{2}}$
with the Fermi momentum $k_{\rm F}$ chosen to give the same number of states,
$N_{\rm s}=k^{2}_{\rm F}/(2\pi) = (2N+1)/(2\pi \ell^{2})$, 
as in the $B\not=0$ case.
The observable vacuum energy density 
$\epsilon_{\rm vac}=\epsilon_{\rm vac}^{B} - \epsilon_{\rm vac}^{B=0}$ 
thereby becomes finite for $N\rightarrow \infty$,
\begin{eqnarray}
\epsilon_{\rm vac}
&=& {\omega_{\rm c}\over{2\pi \ell^{2}}}\, 
\Big[ -\sum_{n=1}^{N} \sqrt{n + \lambda} 
- \textstyle{1\over{2}}\, \sqrt{\lambda} 
\nonumber\\
&& -\textstyle{2\over{3}}\, \lambda^{3/2}  + \textstyle{2\over{3}}\, 
(N + \textstyle{1\over{2}} + \lambda)^{3/2} 
\Big],  \nonumber\\
&=& {\omega_{\rm c}\over{2\pi \ell^{2}}}\, [ -\zeta
(-\textstyle{1\over{2}}) - \textstyle{1\over{2}}\, \sqrt{\lambda} 
+ \cdots ], 
\end{eqnarray}
where $\lambda = {1\over{2}}\, m^{2}\ell^{2}$ and 
$-\zeta(-\textstyle{1\over{2}}) \approx 0.2079$.
The $\chi$ sector also leads to the same $\epsilon_{\rm vac}$.

\newpage



\begin{thebibliography}{99}
\bibitem{fnReg} As in the Tomonaga-Luttinger model, the infinite depth
of the Dirac sea is an artifact of the continuum model (2.1) whose
validity  is limited  to  a (far) smaller energy range than the original
lattice model.   It is this energy range that sets the scale of the
spectrum cutoff, and regularization provides a controlled way to derive
the (low-energy) response independent of the cutoff.
\bibitem{fnRes} To be precise, it is the real part of the response 
(such as $V_{n}^{(2)}$ that enjoys this property; obviously, it is not
shared by the imaginary part.
\bibitem{fnt} To be precise, $\eta_{-N}$ becomes insensitive to 
the transitions across the Dirac sea when the four possible types of
transitions $(-N \rightarrow N \pm 1)$ and $(-N \rightarrow N\pm 2)$ 
are combined.
\end{thebibliography}
\end{document}